# Load-balanced Service Function Chaining in Edge Computing over FiWi Access Networks for Internet of Things

Jing Liu, Guochu Shou, Qingtian Wang, Yaqiong Liu , Yihong Hu, and Zhigang Guo

*Abstract*—Service function chaining (SFC) is promising to implement flexible and scalable virtual network infrastructure for the Internet of Things (IoT). Edge computing is envisioned to be an effective solution to process huge amount of IoT application data. In order to uniformly provide services to IoT applications among the distributed edge computing nodes (ECNs), we present a unified SFC orchestration framework based on the coordination of SDN and NFV, which provides a synergic edge cloud platform by exploiting the connectivity of FiWi access networks. In addition, we study the VNF deployment problem under our synergic framework, and we formulate it as a mixed-integer nonlinear programming (MINLP) problem jointly considering the load balancing of networking and computing for chaining VNFs. We also propose two approximation optimal deployment algorithms named Greedy-Bisection Multi-Path (GBMP) and KSP Multi-Path (KSMP) taking advantage of the multi-instance virtual network functions (VNFs) deployed in ECNs and the multipath capacity in FiWi access networks. Extensive simulations are conducted in two types of IoT application scenarios in the EC over FiWi access networks. The numerical results show that our proposed algorithms are superior to single path and ECMP based deployment algorithms in terms of load balancing, service acceptance ratio, and network utilization in both two typical scenarios.

## I. Introduction

WITH the rapid development of the Internet of Things (IoT), lots of new applications have emerged increasing data traffic and computing resource consumption, such as medical health, smart grid, smart home, AR or VR, intelligent transportation systems and industrial automation [1]. The predominant strategy is to process the data generated by IoT applications in the centralized cloud, which may cause the access and core network congestion and shortage of the computing resource in the data center. Edge computing (EC) is envisioned to be an effective solution to migrate the computing capability from remote cloud data centers to edge computing nodes (ECNs) which are in close to IoT devices (such as sensors, actuators and wearable devices) [2]. So that the IoT applications are able to be offloaded to the ECNs providing computing and storage capacities to the resource limited IoT devices. The representative EC paradigms are fog computing [3], cloudlet [4] and mobile edge computing [5].

The recent trend in IoT is to leverage the service function chain (SFC) customized service to meet the diverse IoT application requirements for deployment and scalability. SFC is defined as a series of service functions (SFs) in a specific order that handle various services/applications traffic, according to the service requirements (e.g., delay sensitive or high bandwidth) and the operators' policies [6]. With SDN [7] and NFV [8] providing virtualization of networks and network functions (SFs in SFC terminology) in cloud computing infrastructures, the IoT application providers can make use of SFC service to control and steer the data flows flexibly traversing a set of virtual network function (VNF) instances (such as firewall (FW), deep packet inspection (DPI), network address translation (NAT), etc.) in a predetermined order, making the IoT application deployment more efficient and scalable [9][10].

A number of research on SFC orchestration and deployment for IoT applications have been emerged over the past two years [11-14]. For instance, Wang *et al.* [11] implemented a cloud SFC orchestration system called PRSFC-IoT under the IoT scenarios, and formulated an ILP optimization problem of performance and resources aware SFC orchestration with an approximation optimization algorithm as a solution. However, most existing research on SFC deployment of IoT applications only focused on the VNFs placement in the centralized cloud, and did not consider the characteristics of networking and computing resource limitation of edge computing. A few works on deploying SFC in edge computing environment for security, energy efficiency and video steaming have been studied recently [15-18]. EC and SFC are seemed as potential technologies to enable the IoT services efficiently deployment. It is promising to integrate the IoT application provider, virtual network infrastructure operator, edge cloud provider, and centralized cloud to provide customized services with edge

The paper is submitted for review on June 15th, 2020. This work was supported by the National Natural Science Foundation of China under Grant 61471053 and Beijing Laboratory of Advanced Information Networks.

J. Liu, G. Shou, Q. Wang, Y. Liu, Y. Hu and Z. Guo are with the Beijing Key Laboratory of Network System Architecture and Convergence, School of Information and Communications Engineering, Beijing University of Posts and Telecommunications, Beijing 100876, China (e-mail: liujing-0115@bupt.edu.cn; gcshou@bupt.edu.cn; qtwang@bupt.edu.cn; liuyaqiong@bupt.edu.cn; yhhu@bupt.edu.cn; gzgang@bupt.edu.cn).

processing and storage capabilities for massive IoT data flows.

We note that there are still several issues of SFC deployment in current edge computing environment for processing IoT applications. Firstly, most of the existing SFC orchestration frameworks are oriented towards centralized cloud in data centers or independently orchestrate computing services on each distributed ECN without considering the cooperation among the VNFs on different ECNs. Specifically, if any ECN or link in the VNF forwarding graph (VNF-FG) [8] of the service chain is failed or overloaded, it will result in the entire SFC failure. Secondly, it is impossible for a single ECN to process plenty of tasks at the same time due to the limitation of computing and storage capabilities. And the IoT devices are geographically widely located in the networks, and the application requests are time-varying and unevenly distributed. This may cause computing resources of some ECNs run out while that of other ECNs are idle, leading to unnecessary services rejection. Therefore, how to dynamically deploy the VNFs and select the forwarding path to achieve load balancing is the key problem.

There have been several works on collaborative edge computing for resource management and orchestration [19], computation tasks offloading [20] or sustainable virtual network embedding in CoTs [21]. These existing works focused on the collaboration among the MEC devices, which did not consider that the cooperation of the VNFs hosted in the ECNs.

In order to address above issues, we consider to design a framework in a service-oriented architecture using network interconnections to turn the standalone service deployment in ECNs into a synergistic SFC orchestration of IoT applications. The FiWi access network which complements the flexibility of wireless access with the high bandwidth of optical access is a potential solution to integrate with EC [22]. We proposed a virtualized integrated networking scheme of multi-access edge computing and FiWi access networks, and the developed testbed verified the effectiveness and performance improvement of the convergence of EC and FiWi access networks [23]. In this paper we propose a unified SFC orchestration framework in collaborative edge computing through the connectivity of the FiWi access networks, which leverages SDN and NFV to provide flexible IoT applications deployment. In addition, we study the VNF deployment problem taking the resource and network topology characteristics of EC over FiWi access networks into consideration to achieve load balancing for different IoT application scenarios. The main contributions of this paper are as follows.

1) We present a synergistic SFC orchestration framework for coordination of all the VNFs on the distributed ECNs in EC over FiWi access networks for IoT applications.
2) We study a VNF deployment problem under the SFC orchestration framework, and formulate it as a mixed-integer nonlinear programming (MINLP) problem, with the objective of the load balancing of the FiWi access networks and ECNs.
3) We propose the greedy-bisection and KSP based approximation optimization deployment algorithms to reduce the complexity of the optimal search strategy. The proposed algorithms achieve better performance in terms of service acceptance ratio, load balancing, and network utilization compared with typical single path and ECMP based deployment algorithms.

The remainder of this paper is organized as follows. The section II introduces the related work on EC and SFC. In section III we propose the SFC orchestration framework in EC over FiWi access networks. Section IV presents the VNF deployment problem formulation. Section V describes our solution. Section VI presents simulations and the result analysis, and section VII summarizes the paper.

## II. RELATED WORK

In this section, we introduce related works on SFC deployment in edge computing, MEC and FiWi access networks, and load balancing in VNFs deployment.

Several SFC deployment and VNF placement methods in edge computing are researched in current works. Blanco *et al.* [15] studied the VNF placement problem for a multitenant cluster of small cells that provide mobile edge computing services. Moreover, they developed a power-aware model and an energy-aware placement solution using a robust optimization approach. The numerical results indicated that the proposal efficiently placed the VNFs in the available hardware infrastructure while fulfilling service constraints. Li *et al.* [16] presented a real-time changing security service chain method for MEC, and a mechanism based on fuzzy theory is proposed to promote the development of EC. The results proved that the proposed mechanism achieved an improved performance in terms of Inverted Generational Distance values. Sun *et al.* [17] proposed a workflow-like service request and a dynamic minimum response time considering same level to map the workflow-like requests in EC. The simulation results demonstrated that the proposed system outperformed in terms of response time delay, blocking rate, and deploy time behavior. The above works have not described the SFC orchestration in edge computing.

There are some existing works related to MEC integrating with FiWi access networks. Maier *et al.* [24] proposed an integrated network architecture and enhanced resource management of MEC over FiWi networks in three typical 5G scenarios, and the obtained results showed the significant benefits in terms of delay, response time efficiency, and battery life of edge devices. To make the best use of the FiWi access networks, Liu *et al.* [25] further proposed a connectivity enhancement scheme and an ECN selection algorithm in the convergence of EC and FiWi access networks. So that different ECNs are able to interconnect with each other only within the access networks. The simulation results demonstrated that the scheme got low delay of tasks offloading and the better load balancing performance.

There are more and more works studying the deployment of SFC to achieve load balancing in different network environments. Kuo *et al.* [26] aimed to solve the joint problem of VNF placement and path selection to make better use of the network. They proposed a systematic approach to adjust the link and server usage for each requirement based on network state and demand attributes. The simulations showed that the design effectively adapted resource usage to network dynamics, and served more demands than other heuristics. Thai *et al.* [27] studied a Nearest First and Local-Global Transformation algorithm, which jointly supported server and network load balancing for chaining VNFs in data center environment. The numerical results indicated that the algorithm increased the system bandwidth utilization up to 45%. Chien *et al.* [29] proposed a service-oriented SDN-SFC load balance mechanism which classified the type and priority of service. And they adopted the heuristic algorithm to plan the transmission paths among SFCs to reduce the load of each SF. The simulation results indicated that the proposed method can shorten the time of data transmission and achieve load balance.

The aforementioned authors do not consider the coordination among the VNFs while deployed in the distributed edge computing environment. To this end, in this paper we jointly consider the load balanced VNF deployment problem and the unified SFC orchestration in distributed EC by leveraging FiWi access networks.

III. SFC ORCHESTRATION FRAMEWORK IN EDGE COMPUTING OVER FIWI ACCESS NETWORKS

The IoT services consist of edge computing applications which belong to cloud services and network services (NSs) that are also known as SFCs. For example, an environment monitoring service includes video caching and analysis as edge computing applications running on ECNs and an SFC consisting of a Firewall and a DPI. The SFC is responsible for connecting users and the edge computing applications. The cloud services deployed by NFV do not take the network connecting requirements between VNFs into consideration. Therefore, centralized control and steer of service flows in network enabled by SDN are necessary. To support the management and deployment of IoT services, we design an SFC orchestration framework loosely coupling with the NFV-MANO. The proposed framework focuses on bringing distributed ECNs into a synergic edge cloud platform by exploiting the connectivity of FiWi access networks and on the combination of SDN, NFV, and cloud to facilitate the provision of edge applications and network services jointly.

*A. Framework design*

Fig. 1 depicts the edge computing and FiWi access networking scheme of SFC orchestration for IoT services includes an orchestration and management layer and a physical infrastructure layer. The orchestration and management layer is responsible for the SFC orchestration, including several functional modules for VNF placement, selecting forwarding paths, scheduling the network, computing and storage resources, and issuing messages to the lower layer. The physical infrastructure layer receives and executes a series of commands to implement data flow forwarding and VNF deployment.

*1) Orchestration and management layer*

The main functional modules in orchestration and management layer are the SFC orchestrator (SFCO), edge computing orchestrator (ECO), edge application manager (EAM), service requirement module, monitoring module, connectivity manager, and placement module. The functions are specified as follows.

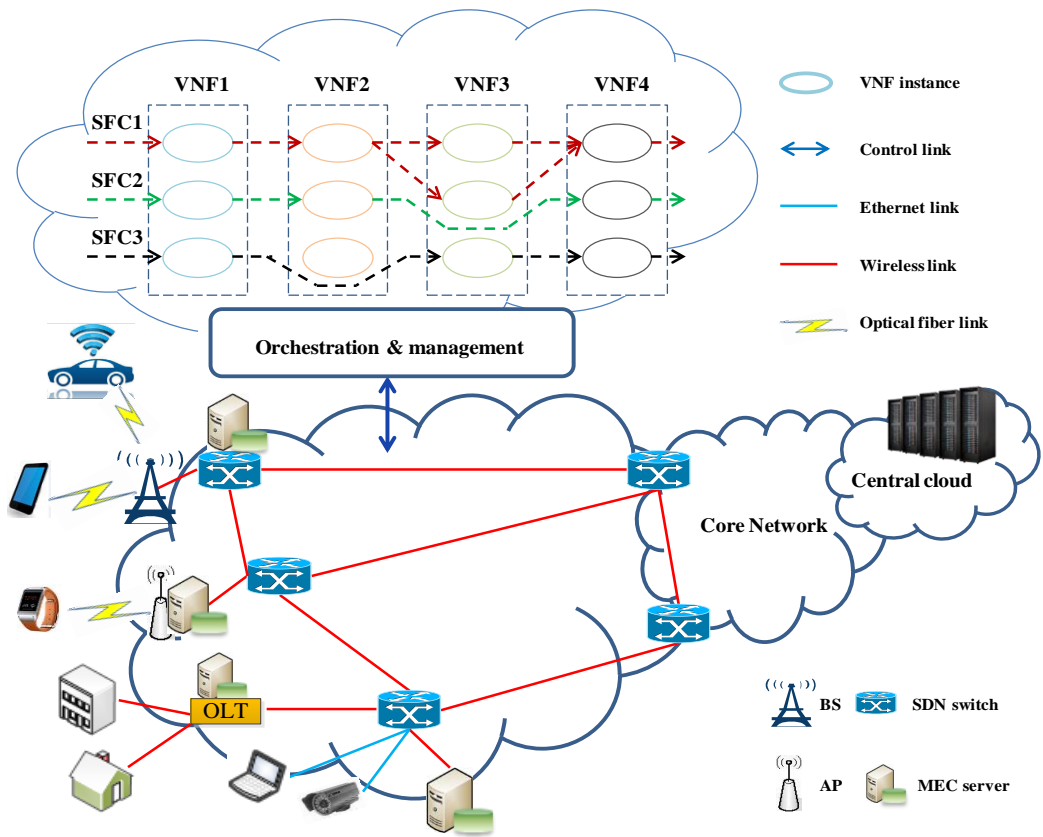

Fig. 1. Illustration of edge computing and FiWi access networking scheme of SFC orchestration for IoT services.

**SFC orchestrator**: is responsible for interacting with ECO and NFV orchestrator. It can orchestrate the horizontal collaboration with different MANOs and ECOs in heterogeneous cloud environments. SFCO receives user requests of edge applications and NSs and dispatches the resource requirements separately to ECO and MANO. The NS is specialized by network service descriptor (NSD) in NFV-MANO [6], which is different from the edge computing applications, hence a unified description model is needed. The Topology and Orchestration Specification for Cloud Applications (TOSCA) standard data model [32] should be extended with new components to enable SFCs to define an edge computing service descriptor [33].

**Edge computing orchestrator (ECO)**: uniformly schedules the resources of all the ECNs managed by multiple VIMs such as OpenStack Heat or Cloudify. It allocates resources to the VNF instances once a service request comes and dynamically launches, configures, and scales up/down the VNFs in all the ECNs. ECO is in charge of the life cycle (such as initiation, deletion, update, and so on) of the edge computing applications through the API. ECO coordinates with SDN controllers like OpenOayLight and ONOS. The SDN controller flexibly steers the service data flow traversing VNFs and edge computing applications with a specific order. ECO is able to provide interconnection among all the ECNs becoming a synergic edge cloud platform through FiWi access networks.

**Edge application manager (EAM)**: provides lifecycle, scaling and fault management for edge computing applications (ECAs). EAM invokes VIM to instantiate the ECA according to the user requirements. It configures the ECA instances and sets the parameters for bootstrap process.

**Service requirement module**: uploads the data size, required bandwidth, and delay to the database. And it is responsible for translating the user requests into appropriate template description of VIM.

**Monitoring module:** monitors the available resources of the network (the available link bandwidth), NFVI (the CPU and memory capacity) and network topology, and updates the information in time. It also monitors the status of the ECA instances in case of failure or overloaded.

**Connectivity manager**: calculates the forwarding routing according to the multi-instance deployment algorithm and selects the link flow table with the optimal factors such as delay and load balance. For example, the response time requirement, the closer the user and the VNF, the shorter the response time and the VNF processing performance. The physical resource correspondence also corresponds to the user waiting time.

**Placement module**: is responsible for decisions on the placement of ECAs in ECNs and NSs to using efficiently the edge computing and FiWi access networks.

*2) Physical infrastructure layer*

The physical infrastructure layer includes a FiWi access network and edge computing nodes. The FiWi access network consists of front-end wireless subnets (including base stations of 4G or 5G mobile communication systems, APs in WLANs with IEEE 802.11g/n/ac/ad, etc.) and back-end fiber subnets (such as EPON and GPON backhaul Net, etc.). The edge computing server is deployed near the base station or AP, or placed near the optical switch or OLT.

### B. SFC Instantiation Procedure

The service deployment procedure is as follows. When the service request arrives, the service requirement collection module outputs the resource requirements of the VNFs and the virtual networks. Taking the topology and the available resource information into consideration, the SFCO compares the load balancing indexes according to the deployment algorithms, and then allocates resources for the VNFs, and creates a virtual multipath network.

For a series of VNFs in an SFC, according to the requirements of computing, storage, and bandwidth, the ECO expands the VNF into multiple parallel instances of the same type if the requirement is greater than a given threshold, and ensures the threshold is not exceeded. The factors such as the number, capacity, and location selection of the VNF extension are obtained according to the service deployment algorithms.

There is a demand to create, configure, and manage these VNFs effectively for a specific IoT application delivered in the edge cloud. To enable such a vision, a resource orchestrating scheme based on the NFV and SDN integrating the distributed ECNs into an edge cloud platform is proposed to fulfill the offloaded tasks from the mobile IoT devices. The edge cloud platform configures the launched VMs to install and perform application-specific tasks. SDN is suitable to work with NFV in the edge cloud converged with FiWi access networks, and configures, controls, and manages the VNFs.

### IV. SYSTEM MODEL AND PROBLEM FORMULATION

We divide the load balancing indicators into three parts: networks, edge computing nodes, and switching nodes. The load balancing problem of SFCs in edge computing over the FiWi access network is formulated as follows.

### A. System Model

Let undirected graph $G = \{N, E\}$ denote the physical network, and *N* represents the nodes in the network, including edge computing servers $N_c$ and switching nodes (switches or FiWi nodes) $N_s$, $N = N_c \cup N_s$, which are indexed by *k*. $C_{ck}^{cpu}$ is used to represent the computing capacity of the computing node $n_{ck}$, $W_{ck}^{cpu}$ is the current computing load of $n_{ck}$. $C_{sk}^{pt}$ indicates the switching port capacity of the switching node $n_{sk}$ (such as the flow table capacity of the OpenFlow switch), $W_{sk}^{pt}$ indicates the current traffic load of $n_{sk}$. Let *E* indicate the links in the network, including the optical link $E_o$ and the wireless link $E_w$, $E = E_o \cup E_w$, indexed by *l*. The bandwidth of the optical link $e_{ol}$ and the wireless link $e_{wl}$ are respectively represented by $B_{ol}$ and $B_{wl}$, and $W_{ol}^{bw}$ and $W_{wl}^{bw}$ represent the current link traffic load of $e_{ol}$ and respectively.

Let $S$ represents the service chain requests, which is indexed by $i, 1 \leq i \leq N$, and $s_i$ denotes the $i$th service chain request in the network. Each service chain request $s_i$ includes an ingress node $O_i$, an egress node $T_i$, the order of a series of VNFs, computing capacity demand $D_i^{cpu}$ of the VNF, link bandwidth demand $D_i^{bw}$ of the service flows, and the longest tolerable end-to-end delay $D_i^{delay}$ of the service.

The service chain $s_i$ is implemented by $m$ kinds of VNFs, which is represented by $F_i = \{f_1, f_2, \ldots, f_m\}$, and $|F_i|$ indicates the length of $s_i$, and let $j$ represent the number of instances of the same type of VNF existing simultaneously, and $f_{mj}^i$ represents the $j$th instance of the $m$th VNF through which $s_i$ traverses. The shortest distance between the edge computing nodes which carry the VNF instance from $f_{mj}^i$ to $f_{m+1,j'}^i$ is denoted by $h_{mjj'}^i$, expressed by the hop counts. We assume that multiple instances of a VNF can be deployed on different edge computing nodes, and each VNF instance can only be hosted by one edge computing node. An edge computing node can carry multiple different VNFs, and if the computing capacity of the VNF is large enough, it can handle different service chain requests.

Let the undirected graph $\bar{G}_i = \{\bar{N}_i, \bar{E}_i\}$ denote the virtual network satisfying the demand of the service chain $s_i$, and $\bar{N}_i$ denotes the virtual node set required to deploy $F_i$, and $\bar{E}_i$ denotes the virtual link set that flow of $s_i$ traverses the specific order VNF instances from $O_i$ to $T_i$. Then $\bar{G}_i$ is embedded in the physical network $G$ that satisfies the requirements. $\bar{N}_i$ is mapped to the physical nodes $N$, including the edge computing nodes and the switching nodes, and a VNF of one type is used for multi-instance which can be hosted on multiple edge computing nodes. $\bar{E}_i$ is mapped to the links connecting all the selected physical nodes in the physical network, and the path from the ingress to the egress node can be multipath.

Load balancing includes load balancing of networks, edge computing nodes, and switching nodes. We define an imbalance indicator $LBI = \{LBI_c, LBI_n, LBI_s\}$ to indicate the degree of load balancing, respectively.

The $LBI_c$ of the ECNs is shown in (1). The numerator is the largest load among all edge computing nodes, and the denominator is the average load of all edge computing nodes in the network.

$$LBI_c = \frac{Max_{i \in M}\{\sum_{j=1}^{N} W_j\}}{\frac{1}{M}\sum_{i=1}^{M}\sum_{j=1}^{N} W_j} \quad (1)$$

The network $LBI_n$ is defined as the ratio of maximum bandwidth utilization to the average bandwidth utilization. In particular, the bandwidth utilization is the ratio of the traffic of all VNF instances on each link to the total bandwidth in the network, as in (2).

$$LBI_n = \sqrt{\frac{1}{L}\sum_i \sum_{l=1}^{L}\left(\frac{W}{B} - \frac{W}{BL}\right)^2} \quad (2)$$

The $LBI_s$ of the switching nodes is defined as the maximum bandwidth utilization of the traffic of all VNF instances on each path in the network through each link, as in (3).

$$LBI_s = \frac{Max_{i \in M}\{\sum_{j=1}^{N} C_j\}}{\frac{1}{M}\sum_{i=1}^{M}\sum_{j=1}^{N} C_j} \quad (3)$$

*B. Problem Formulation*

Let the binary variable $x_{mjk}^i$ denote whether the $j$th instance $f_{m,j}^i$ of the $m$th VNF of the $i$th service chain $s_i$ is deployed on the edge computing node $n_{ck}$.

Let $\xi_{mj,j'}^i$ denote the distribution ratio of the amount of computing and flow traffic between instances from $f_{m,j}^i$ to $f_{m+1,j'}^i$ of two VNFs which are in adjacent order of the service chain $s_i$.

The objective function is to minimize the load balancing indicators of the network link bandwidth utilization, edge computing nodes and switching nodes.

$$\min(\alpha LBI_c + \beta LBI_n + \gamma LBI_s) \quad (4)$$
$$s.t.\ C1:\ \sum_{j=1}^{N} x_{mjk}^i W_j \leq C \quad (5)$$
$$C2:\ \sum_{i=1}^{N} b\, x_{mjk}^i \leq B \quad (6)$$
$$C3:\ \sum_{i=1}^{N}(t_{exe} + t_T) \leq D_i^{delay} \quad (7)$$
$$C4:\ x_{mjk}^i \in \{0,1\} \quad (8)$$

The constraints *C1* and *C2* are that the sum of all the sub-flows through a link is not greater than the link bandwidth. The resources occupied by all VNF instances on one edge computing server are not greater than the capacity of the computing capacity. The constraint *C3* ensures that the delay of an SFC is less than its maximum delay the flow can tolerate. The constraint *C4* ensures *the VNF deployment variable is integer.*

## V. SOLUTION

In this section we propose two service deployment algorithms to achieve ECNs, networks and switching load balancing. The first concern is the VNF multi-instance deploying model. In other words, one type of VNFs has more than one instance which are hosted in different ECNs. The second concern is the multipath capacity of FiWi access networks. That is, SFC flows transmitting between VNFs can go through multiple paths. A VNF that has multiple instances simultaneously provides the possibility to route the data flow through multipath. It is challengeable to load-balanced deploy the multi-instance VNFs on distributed ECNs in the FiWi access networks with abundant multipath capacity.

*A. Greedy-Bisection Multipath Algorithm (GBMP)*

In the GBMP algorithm, the greedy-bisection algorithm is firstly applied to enumerate the alternative path sets that can be deployed, and then the linear programming is used to calculate the allocation ratio of the data computing amount and the bandwidth occupation of the adjacent sequential VNF instances, and then the multipath to achieve the minimum load balancing indicator is selected and deployed. We use local optimum to approximate global optimality. The pseudo code is as shown in Algorithm 1.

$n_s$ and $n_d$ represent the original node and the terminal node of the service flow, respectively. $\Xi_i$ represents the computing amount of $s_i$ and the allocation of traffic. $P_i$ represents the deployment path of $s_i$, and $\Lambda_i$ represents the deployment plan. $MP$ represents the maximum number of multipath.

```
Algorithm 1: GBMP
Input: $s_i$, $\omega_i$, $G$, $W_k$, $n_s$, $n_d$, $R$, $MP$
Output: $\Xi_i$, $P_i$, $\Lambda_i$
 1:  Deploy NF $f_m^i$ in $s_i$
 2:  for $m=1 : M$
 3:    Find the alternative node set $N_{Am}^i$ $\{\max W_k, size(N_{Am}^i) \leq MP\}$, and sort the set descending order.
 4:    for $j'=1 : size(N_{Am}^i)$
 5:      if $\left(u_{N_{Amj'}^i} + \omega_i\right) \leq mean \sum \omega_i$
 6:        Deploy $f_m^i$ on $N_{Amj'}^i$
 7:        Break
 8:      else
 9:        Formulated as a min-max problem according to (5).
10:        Solve with greedy-bisection algorithm.
11:      end if
12:    end for
13:    Update $\Xi_i$, $P_i$ and $\Lambda_i$ with $\xi_{0j,j'}^i$
14:    Update $R_{mj'}^i$ and $b_{mj,j'}^i$
15:  end for
16:  Return $\Xi, \Lambda, P$
```

For the location selection of the VNFs of each service function chain, we use the node degree $d_k$ and the betweeness centrality $BC_k$ of the node $n_k$ to indicate the importance of the location of the node in the network. $dist_{mjk}$ represents the distance from node $n_k$ to the last hop node $n_{mj}$, expressed in hop counts. $b_{mjk}$ represents the available bandwidth of the node $n_k$ o the last hop node $n_{mj}$. Let $W_k$ denote the weight of the node $n_k$, which is represented by the sum of the intermediately and the degree of the node and the normalized ratio of the available bandwidth and the shortest distance between the two nodes. The greater the weight, the higher priority of the node is. The weight is as shown in (9).

$$W_k = (BC_k + d_k)/b_{mjk} dist_{mjk} \tag{9}$$

$$min \ Max\{\frac{N}{\sum_i \omega_i}\sum_{j'}\sum_j(\xi_{mj,j'}^i + u_{N_{Am+1j'}^i})\} \tag{10}$$

To deploy the VNFs, we need to select a set of available alternative nodes with the highest node weight, where the number of nodes is smaller than *MP* and sorted in descending order. When deploying the first NF, let *m=0*, i.e. $n_{mj}=n_s$. Then comparing whether the available resources of the first element of the set are greater than the average load of all edge computing nodes in the network. If it is greater than the average load, the unique instance of $f_m^i$ will be deployed on the node; otherwise, according to (8). The maximum value problem is used to construct the objective function, and the greedy-bisection algorithm is used to solve the optimal deployment node. Then update the remaining resources of these nodes and the available link bandwidth.

*B. K-shortest-path Multipath Algorithm (KSMP)*

The above algorithm converts the objective function into a

```
Algorithm 2: KSMP
Input: $s_i$, $b_i$, $G$, $W_k$, $n_s$, $n_d$, $R$, $MP$
Output: $\Xi_i$, $P_i$, $\Lambda_i$
 1:  Deploy NF $f_m^i$ in $s_i$
 2:  for $m=1 : M$
 3:    if $m=1$
 4:      Let
              $cost_k = (B_{sk} - b_{sk}) + u_k$
 5:    else
 6:      Let
              $cost_k = (B_{m-1,jk} - b_{m-1,jk}) + u_k$
 7:    end if
 8:    for $iter = 1: MP$
 9:      $[P_{Am}^i, Cost_{Am}^i]$=KSP $(G, n_{m-1,j}, n_k, iter)$
10:      if $max(K) < iter$
11:        break
12:      end if
13:    end for
14:    for $j'=1 : size(Cost_{Am}^i)$
15:      Formulated as a LP problem according to (6).
16:      Solve with Simplex algorithm
17:    end for
18:    Return $\Xi_i$, $P_i$, $\Lambda_i$ with $\xi_{mj,j'}^i$
19:    Return $R_{mj'}^i$ and $b_{mj,j'}^i$
20:  end for
21:  Return $\Xi, \Lambda, P$
```

problem of optimizing the server load with the link bandwidth as a constraint. In the KSMP, we take the server and network load as the path cost and regard the deployment problem as a multipath selection problem. The k-shortest-path algorithm is a solution of the objective function. The pseudo code is as shown in Algorithm 2.

When the $m^{th}$ NF $f_m^i$ of the service chain $s_i$ is deployed, the sum of the load of the path from the nodes hosting all instances of the $(m-1)^{th}$ NF $f_{m-1,j}^i$ to the next hop node $n_k$ and the load of the node $n_k$, are regarded as the path cost, respectively. We define $f_{m-1,j}^i$ as $n_s$ when deploying the first NF $f_1^i$. Then, the KSP algorithm is invoked to sequentially find the set of 1 to MP paths and the cost set of the corresponding path in order of path cost from small to large. If the maximum number of paths that can be found is less than *MP*, stop searching. Then, according to (11), the objective function is transformed into a minimum linear programming problem, and the simplex algorithm is used to find the minimum cost splitting ratio between two NFs in each adjacent node pairs. If the resources of all nodes or links in the network cannot meet the requirements of $s_i$, the service request fails.

The constraint indicates that the sum of all offloading traffic on each link should be less than or equal to the link bandwidth, and all processing workload of each edge computing node should be less than or equal to its computing resources.

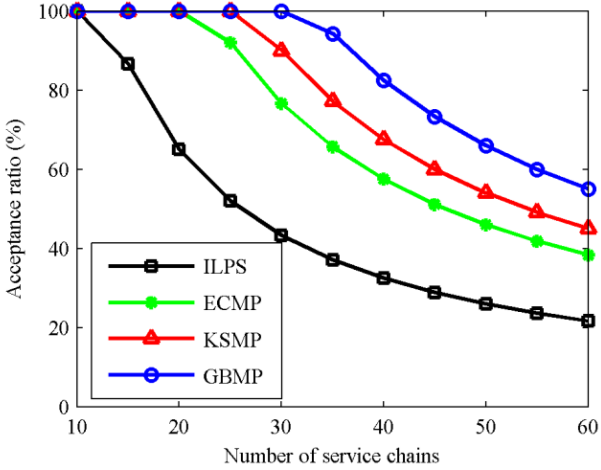

Fig. 2. Service request acceptance ratio vs. number of service chains.

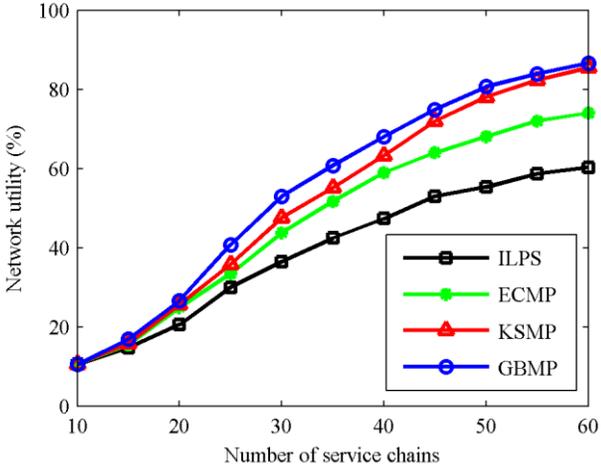

Fig. 3. Network utilization vs. number of service chains.

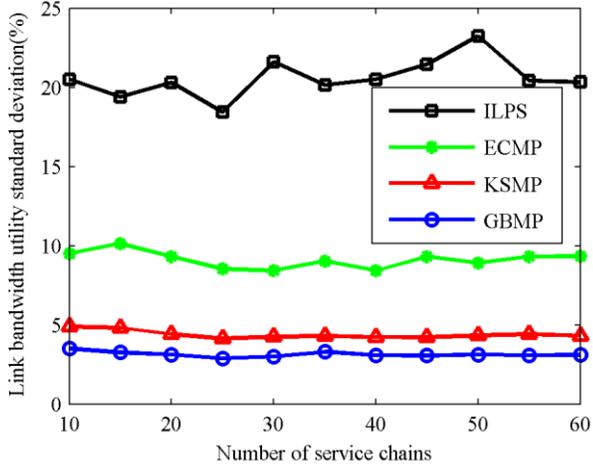

Fig. 4. Standard deviations of link bandwidth utilization vs. number of service chains.

$$min \sum_{j=1}^{J} \sum_{j'=1}^{J'} \xi_{m j,j'}^{i} cost_{jj'} \quad (11)$$
$$s.t. \ (5)\text{-}(7)$$

## VI. PERFORMANCE EVALUATION

The network topology is set up in a tree-to-star structure. There are 15 ECNs, and the ECNs interconnect with each other via fiber links. The VNF is hosted in the ECN, and the service traffic arrives the VNF instances through the multipath of the optical network forwarded by the switching nodes. The computing capacity of each ECN in the network is a uniform distribution with a mean value of 30 Gigacycles, a variance of 31.8. The parameters about FiWi access networks are such as the optical link bandwidth of 10 Gbps, the channel bandwidth of 54 Mbps per wireless link, and a FiWi node with 4 channels. The available computing capacity of the ECN and the available bandwidth of the link are evenly distributed. We use MATLAB for simulation.

The IoT applications are divided into two service types, i.e., data-intensive and user-intensive, according to the amount of collected data, number of requests generated concurrently and demands of computing resource of the applications.

### A. Data-intensive Service Scenario

Examples of data-intensive applications include AR or VR applications using Oculus Rift helmets and Google Glasses. The computing and storage resources can also be assigned and utilized more efficiently.

The number of VNF types for each service request is evenly distributed between 3 and 5. The CPU resource requirements generated by each service vary from 500 Megacycles to 1 Gigacycles, and the amount of service data is subject to a uniform distribution with an average of 800 MB and a variance of 4. The number of requests for the service chain has gradually increased from 10 to 60.

We compare the KSMP algorithm and the GBMP algorithm with the typical equal cost multipath algorithm (ECMP) and the single-path linear programming (ILPS) deployment algorithm in the convergence of FiWi access networks and edge cloud scenarios. The server load balancing indicators, link load balancing, network utilization, and service acceptance ratio in the entire network are used as four comparison performance metrics of the above algorithms. We use GBMP to represent multipath greedy-bisection algorithm, KSMP to represent KSP multipath algorithm, ECMP to represent equal cost multipath algorithm and ILPS to represent integer linear programming single path deployment algorithm.

Fig. 2 shows the comparison of the service request acceptance ratio for the four algorithms. The acceptance ratio of ILPS begins to decline at the number of services reaches 10, and the acceptance ratio of KSMP begins to decline when the number of services reaches 28, while GBMP begins to decline when the number of services reaches 34. The acceptance ratio of GBMP is 21.4% higher than that of KSMP, and the acceptance ratio of ILPS is increased by 240%. The GBMP link capacity of multipath and the computing of multi-instance of the GBMP are larger than the single path, and is more load balancing than ECMP.

Fig. 3 illustrates the network utilization of different algorithms. It represents that GBMP gets the highest utilization reaching about 84% as 60 service chain requests arrive. Network utilization of KSMP is slightly lower than GBMP. When the service request is 60, the network utilization of

ECMP is 74%, and the ILPS is the lowest, at 58%.

Fig. 4 shows the comparison of the standard deviations of the link bandwidth utilization of the four algorithms with the increase of service requests for data-intensive services. The larger value of the standard deviation gets, the more unbalanced the link load is. It can be seen from the figure that the GBMP algorithm has the highest load balance performance under our integrated edge cloud platform.

Fig. 5 shows the load balancing indicator comparison of the four algorithms as the service request increases. The larger the indicator, the more unbalanced the load is. It can be seen from the figure that GBMP has the lowest load balancing indicator and the best performance.

### B. User-intensive Service Scenario

We define user-intensive application referring to the wide area IoT use cases consisting of large numbers of low-cost devices (such as WSN) with large scale and high requirements on scalability. Examples of user-intensive applications include structure or agricultural field monitoring, and industrial monitoring, which usually involve a lot of small-sized and distributed sensors and devices. In these cases, the sensors and IoT devices mostly generate and send data but with limited capacity in processing and analyzing data.

The number of VNF types for each service request is still evenly distributed between 3 and 5. The CPU resource requirements generated by each service vary from 10 to 100 Metacycles, and the amount of service data is evenly distributed between 300 and 800 KB. The number of requests for the service chain has gradually increased from 100 to 1,000.

Fig. 6 shows a comparison of the service request acceptance ratio for the four algorithms. The acceptance ratio of ILPS and ECMP turn to decline at the number of services requests of 245 and 333, and the acceptance ratio of KSMP begins to decline when the number of services reaches 389, while GBMP begins to decline when the number of services reaches 421. The acceptance ratio of GBMP is 8.23%, 26.43%, and 71.84% higher than that of KSMP, ECMP and ILPS, respectively.

Fig. 7 shows a comparison of load balancing indicators for four algorithms for user-intensive services as requests increase. It can be seen from the figure that the GBMP algorithm has the highest load balance under our integrated edge cloud platform.

Statistical values of the network utilization and standard deviations of the link bandwidth utilization may vary due to the different scenarios, and we found the trend of graphs more or less the same in each service type, from which performance of our algorithms can be seen that outperform others.

From the above results analysis, we can see that combined with the FiWi access networks and the edge computing simulation environment, the multipath based deployment methods have significant advantages over the single path based deployment methods in terms of multiple indicators. For the multi-instance VNF scenario, the multipath service acceptance ratio is 200%-240% higher than the single path, and the imbalance degree is reduced by 58-80%. Compared with the two multipath methods, the multipath implementation of traditional networks can only be ECMP, the proposed method based on our framework is valid for both service scenarios. The multipath method is more efficient in the data/ user-intensive scenario.

### VII. CONCLUSION

In this paper we propose an SFC orchestration framework based on SDN and NFV, which provides a synergic edge cloud

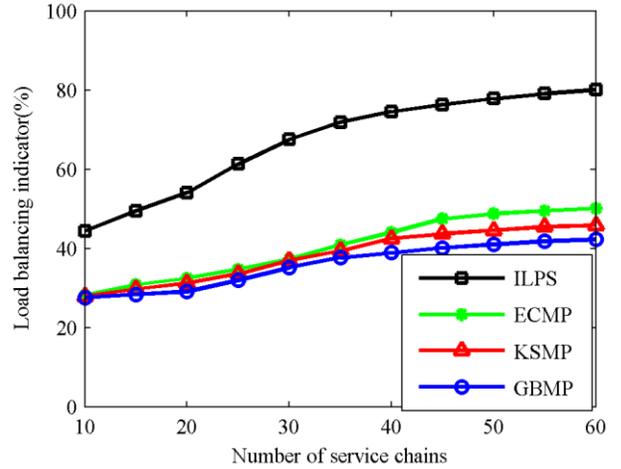
Fig. 5.  Load balancing indicator vs. number of service chains.

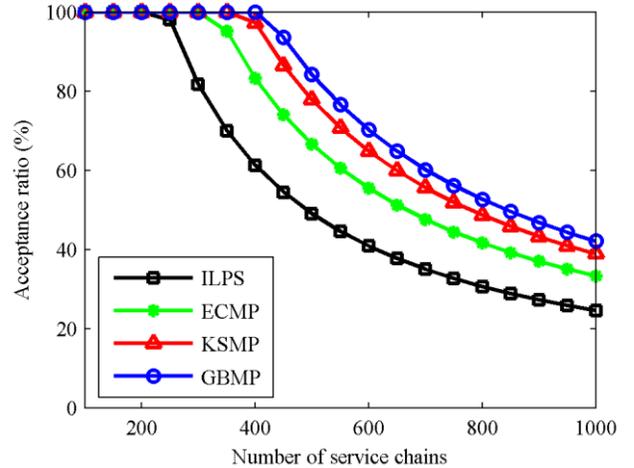
Fig. 6.  Service request acceptance ratio vs. number of service chains.

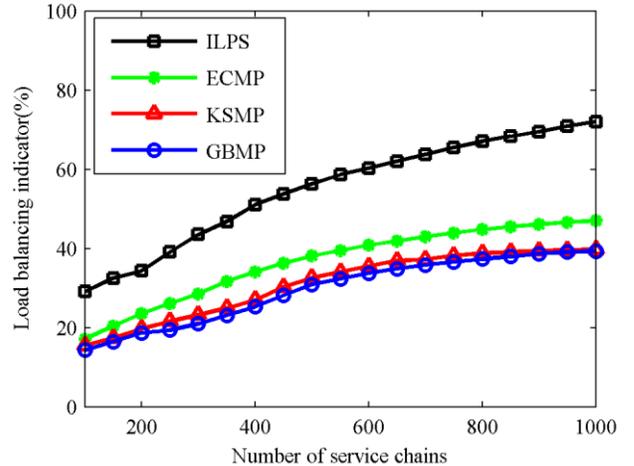
Fig. 7.  Load balancing indicator vs. number of service chains.

platform by leveraging the high bandwidth and flexibility access of FiWi access networks. In addition, we formulate a load balancing problem of both ECNs and FiWi access networks, and design two heuristic algorithms GBMP and KSMP based on greedy-bisection and KSP algorithms respectively. Finally, extensive simulations are conducted in two types of IoT application scenarios, i.e., data-intensive and user-intensive, in the edge computing over FiWi access networks. The numerical results show that our proposed algorithms are superior to single path and ECMP based deployment algorithms in terms of service acceptance ratio, network utilization and load balancing in both two typical application scenarios. Taking the advantage of the convergence of FiWi access networks and edge computing based on SDN and NFV makes the implementation of multipath-based SFC deployment algorithms more efficiently. Therefore, the proposed framework is able to meet the requirements of load balanced SFC deployment for newly emerging edge IoT applications.

REFERENCES

[1] A. Al-Fuqaha, M. Guizani, M. Mohammadi, M. Aledhari, and M. Ayyash, "Internet of Things: A Survey on Enabling Technologies, Protocols, and Applications," *IEEE Commun. Surveys Tuts*, vol. 17, no. 4, pp. 2347-2376, 4th Quart., 2015.
[2] J. Pan and J. McElhannon, "Future Edge Cloud and Edge Computing for Internet of Things Applications," *IEEE Internet Things J.*, vol. 5, no. 1, pp. 439-449, Feb. 2018.
[3] M. Chiang and T. Zhang, "Fog and IoT: An overview of research opportunities," *IEEE Internet Things J.*, vol. 3, no. 6, pp. 854–864, Dec. 2016.
[4] M. Satyanarayanan, P. Bahl, R. Caceres, and N. Davies, "The case for VM-based cloudlets in mobile computing," *IEEE Pervasive Comput.*, vol. 8, no. 4, pp. 14–23, Oct./Dec. 2009.
[5] M. Patel, B. Naughton, and et al., "Mobile-edge computing-introductory technical white paper," ETSI, White Paper 18-09-14, Sep. 2014.
[6] A. M. Medhat, T. Taleb, A. Elmangoush, G. A. Carella, S. Covaci, and T. Magedanz, "Service Function Chaining in Next Generation Networks: State of the Art and Research Challenges," *IEEE Comm. Mag.*, vol. 55, no. 2, pp. 216–223, Feb. 2017.
[7] B. A. A. Nunes, M. Mendonca, X.-N. Nguyen, K. Obraczka, and T. Turletti, "A Survey of Software-defined Networking: Past, present, and future of programmable networks," *IEEE Commun. Surveys Tuts.*, vol. 16, no. 3, pp. 1617–1634, 3rd Quart., 2014.
[8] R. Mijumbi, J. Serrat, et al., "Network function virtualization: State of the-art and research challenges," *IEEE Commun. Surveys Tuts.*, vol. 18, no. 1, pp. 236–262, 2016.
[9] J. Halpern and C. Pignataro, "Service Function Chaining (SFC) Architecture," RFC 7665, Oct. 2015.
[10] Network Functions Virtualisation (NFV); Architectural Framework, document ETSI GS NFV 002 v1.2.1, ETSI NFV, Dec. 2014.
[11] J. Wang, H. Qi, K. Li, and X. Zhou, "PRSFC-IoT: A Performance and Resource Aware Orchestration System of Service Function Chaining for Internet of Things," *IEEE Internet Things J.*, vol. 5, no. 3, pp. 1400-1410, Jun. 2018.
[12] D. Zou, Z. Huang, B. Yuan, H. Chen, and H. Jin, "Solving Anomalies in NFV-SDN Based Service Function Chaining Composition for IoT Network," *IEEE Access*, vol. 6, pp. 62286-62295, Nov. 2018.
[13] R. Kouah, A. Alleg, A. Laraba, and T. Ahmed, "Energy-Aware Placement for IoT-Service Function Chain," in *Proc. IEEE CAMAD.*, Sept. 2018 .
[14] K.-M. KO, A. M. Mansoor, R. Ahmad, and S.-G. Kim, "Efficient Deployment of Service Function Chains (SFCs) in a Self-Organizing SDN-NFV Networking Architecture to Support IOT," in *Proc IEEE ICUFN*, Jul. 2018 .
[15] B. Blanco, I. Taboada, J. O. Fajardo, et al., "A Robust Optimization based Energy-Aware Virtual Network Function Placement Proposal for Small Cell 5G Networks with Mobile Edge Computing Capabilities," in *Proc Mobile Inf. Syst.* , 2017, pp.1–14.
[16] G. Li, H. Zhou, B. Feng, et al., "Fuzzy theory based security service chaining for sustainable mobile-edge computing," in *Proc Mobile Inf. Syst.* , 2017, pp. 1–13.
[17] G. Sun, Y. Li, Y. Li, D. Liao, and V. Chang, "Low-latency Orchestration for Workflow-oriented Service Function Chain in Edge Computing," Future Generation Computer Systems, 2018.
[18] S. Salsano, L. Chiaraviglio, N. B.-Melazzi, et al., "Toward superfluid deployment of virtual functions: exploiting mobile edge computing for video streaming," in *Proc IEEE Int. Teletraffic Cong.*, vol. 2, 2017, pp. 48-53.
[19] T. X. Tran, A. Hajisami, P. Pandey, and D. Pompili, "Collaborative Mobile Edge Computing in 5G Networks: New Paradigms, Scenarios, and Challenges", *IEEE Commun. Mag.*, vol. 55, no. 4, pp. 54-61, Apr. 2017.
[20] H. Guo, J. Liu, and H. Qin, "Collaborative Mobile Edge Computation Offloading for IoT over Fiber-Wireless Networks", *IEEE Network*, vol.32, no. 1, Feb. 2018.
[21] Z. Ning, X. Kong, F. Xia, W. Hou, and X. Wang, "Green and Sustainable Cloud of Things: Enabling Collaborative Edge Computing", *IEEE Commun. Mag.*, vol. 57, no. 1, pp. 72-78, Jan. 2019.
[22] M. Maier and B. P. Rimal, "Invited Paper: The Audacity of Fiber-Wireless (FiWi) Networks: Revisited for Clouds and Cloudlets," *China Commun.*, vol. 12, no. 8, pp. 33-45, Aug. 2015.
[23] J. Liu, G. Shou, Y. Liu, Y. Hu, and Z. Guo, "Performance Evaluation of Integrated Multi-access Edge Computing and Fiber-wireless Access Networks." *IEEE Access*, vol. 6, pp. 30269-30279, May 2018.
[24] B. P. Rimal, D. P. Van, and M. Maier, "Mobile-Edge Computing Empowered Fiber-Wireless Access Networks in the 5G Era," *IEEE Commun. Mag.*, vol. 11, no. 2, pp. 192-200, Nov. 2017.
[25] J. Liu, G. Shou, and J. Xue, "Connectivity Enhancement of Edge Computing Over Fiber-Wireless Access Networks for IoT," in *Proc OSA/IEEE ACP*, Oct. 2018, pp. .
[26] T.-W. Kuo, B.-H. Liou, K. C.-J. Lin, et al., "Deploying chains of virtual network functions: On the relation between link and server usage," in *Proc. IEEE INFOCOM*, 2016, pp. 1-9.
[27] M.-T. Thai, Y.-D. Lin, and Y.-C. Lai, "Joint server and network optimization toward load-balanced service chaining," *Int. J. Commun. Sys.*, vol. 31, no. 10, pp. 1-12, Jul. 2018.
[28] L. Wang, Z. Lu, X. Wen, R. Knopp, and R. Gupta, "Joint Optimization of Service Function Chaining and Resource Allocation in Network Function Virtualization," *IEEE Access*, vol. 4, pp. 8084-8094, Nov. 2016.
[29] W. Chien, C. Lai, H. Cho, and H. Chao, "A SDN-SFC-based Service-Oriented Load Balancing for the IoT Applications," *J. Comp. Netw. App.*, vol. 114, pp. 88-97, Jul. 2018.
[30] W. Ding, W. Qi, J. Wang, and B. Chen, "OpenSCaaS: An Open Service Chain as a Service Platform Toward the Integration of SDN and NFV." *IEEE Network*, vol. 29, no. 3, 2015, pp.30-35.
[31] Q. Wang, G. Shou, Y. Liu, Y. Hu, Z. Guo, and W. Chang, "Implementation of Multipath Network Virtualization With SDN and NFV", *IEEE Access*, vol. 6, pp. 32460-32470, May 2018.
[32] OASIS, "Topology and Orchestration Specification for Cloud Applications, available: http://docs.oasisopen.org/tosca/TOSCA-Simple-Profile-YAML/v1.0/csprd02/TOSCASimple-Profile-YAML-v1.0-csprd02.pdf." 2016.
[33] M. Mechtri, I. G. Benyahia, and D. Zeghlache, "Agile Service Manager for 5G," 2016 IEEE/IFIP NOMS 2016 Istanbul, Turkey, Apr. 25–29, 2016, pp. 1285–90, http://dx.doi. org/10.1109/NOMS.2016.7503004.